\documentclass[aps,prl,twocolumn,superscriptaddress,amsmath,amsfonts,amssymb]{revtex4}

\usepackage{graphicx}
\usepackage{bm}
\usepackage{color}
\usepackage{nicefrac}
\usepackage{amsmath}
\usepackage{verbatim}
\usepackage{longtable}
\usepackage[normalem]{ulem}

\newcommand{\Enull}{\sqrt{\Delta_{\rm so}^2+4\Delta_{KK'}^2}}

\newcommand{\avg}[1]{\left< #1 \right>}
\newcommand{\bra}[1]{\langle #1|}
\newcommand{\ket}[1]{|#1\rangle}

\newcommand{\adag}{a^{\dagger}}

\renewcommand{\vec}[1]{\mathbf{#1}}

\newcommand{\hmprime}{\frac{du}{dz}}

\begin{document}

\title{Spin-Orbit-Induced Strong Coupling of a Single Spin to a Nanomechanical Resonator}

\author{Andr\'as P\'alyi}
\affiliation{Department of Physics, University of Konstanz, D-78457 Konstanz, Germany}
\affiliation{Department of Materials Physics, E\"otv\"os University, 
H-1517 Budapest POB 32, Hungary}

\author{P. R. Struck}
\affiliation{Department of Physics, University of Konstanz, D-78457 Konstanz, Germany}

\author{Mark Rudner}
\affiliation{Department of Physics, Harvard University, Cambridge, Massachusetts 02138, USA}

\author{Karsten Flensberg}
\affiliation{Department of Physics, Harvard University, Cambridge, Massachusetts 02138, USA}
\affiliation{Niels Bohr Institute, University of Copenhagen, Universitetsparken 5, DK-2100 Copenhagen, Denmark}

\author{Guido Burkard} 
\affiliation{Department of Physics, University of Konstanz, D-78457 Konstanz, Germany}

\begin{abstract}
We theoretically investigate the deflection-induced
coupling of an electron spin to vibrational motion due to 
spin-orbit coupling in suspended carbon nanotube quantum dots.
Our estimates indicate that, with current capabilities, a quantum dot with an odd number of electrons can serve as a realization of the Jaynes-Cummings model of quantum electrodynamics in the strong-coupling regime. 
A quantized flexural mode of the suspended tube plays the role of the optical mode and
we identify two distinct two-level subspaces, at small and large magnetic field, which can be used as qubits in this setup.
The strong intrinsic spin-mechanical coupling allows for detection, as well as manipulation of the spin qubit, and may yield enhanced performance of nanotubes in sensing applications.  
\end{abstract}

\maketitle

%
Recent experiments in nanomechanics have reached the ultimate quantum
limit by cooling a nanomechanical system close to its ground state  \cite{OConnell-groundstate}.
Among the variety of available nanomechanical systems, nanostructures
made out of atomically-thin carbon-based materials such as graphene
and carbon nanotubes (CNTs) stand out 
due to their low masses and high stiffnesses.
These properties give rise to high oscillation frequencies, potentially enabling near ground-state cooling using conventional cryogenics, and large zero-point motion, which improves the ease of detection \cite{Peng06,Chen09}. 

\begin{figure}
	\begin{center}
		\includegraphics[width=0.45 \textwidth]{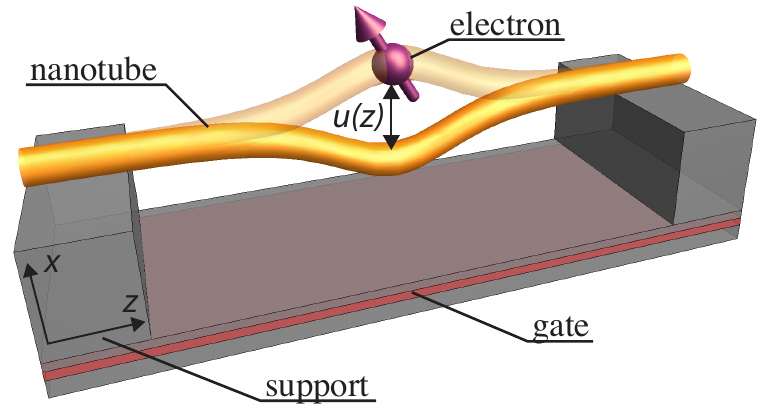}
	\end{center}
	\caption{Schematic of a suspended carbon nanotube (CNT) 
          containing a quantum dot filled with a single electron spin.  The
          spin-orbit coupling in the CNT induces a strong coupling
          between the spin and the quantized mechanical motion
          of the CNT.
	\label{fig:setup}}
\end{figure}

Recently, a high quality-factor suspended CNT resonator was used to demonstrate strong coupling between nanomechanical motion and single-charge tunneling through a quantum dot (QD) defined in the CNT \cite{steele09}.
Here, we theoretically investigate the coupling of a single
electron spin to the quantized motion of a discrete flexural mode of a suspended CNT 
(see Fig.\ref{fig:setup}), and show that the strong-coupling regime of this 
Jaynes-Cummings-type system is within reach.
This coupling provides means for  electrical
manipulation of the electron spin via microwave irradiation, 
and leads to strong nonlinearities in the CNT's mechanical response which may potentially be used for enhanced functionality in sensing applications \cite{Chiu08, Lassagne08, Lassagne11}.

In addition to their outstanding mechanical properties, carbon-based systems also possess many attractive characteristics for information processing applications.
The potential for single electron spins in QDs to 
serve as the elementary qubits for quantum information processing \cite{loss-divincenzo} is currently being investigated in a variety of systems.
In many materials, such as GaAs, the hyperfine interaction between electron and nuclear spins is the primary source of electron spin decoherence which limits qubit performance (see e.g., \cite{Petta}).
However, carbon-based structures can be grown using starting materials isotopically-enriched in $^{12}$C, which has no net nuclear spin, thus practically eliminating the hyperfine mechanism of decoherence \cite{trauzettel07}, leaving behind only a spin-orbit contribution \cite{bulaev08, ps}. 
Furthermore, while the phonon continuum in bulk materials provides the primary bath enabling spin relaxation, the discretized phonon spectrum of a 
suspended CNT can be engineered to have an extremely low density of states at the qubit (spin) energy splitting. 
Thus very long spin lifetimes are expected 
off-resonance \cite{Cottet-fmqubit}. 
On the other hand, when the spin splitting is nearly resonant 
with one of the high-Q discrete phonon ``cavity'' modes, strong
spin-phonon coupling can enable qubit control, information transfer,
or the preparation of entangled states.

The interaction between nanomechanical resonators and single spins was recently 
detected \cite{Rugar-singlespindetection}, and has been theoretically investigated \cite{Rabl-strongcoupling,Rabl-spinphonon} for cases where 
the spin-resonator coupling arises from the relative motion of the spin and a source of local magnetic field gradients.
Such coupling is achieved, e.g., 
using a magnetic tip on a vibrating cantilever which can be positioned
close to an isolated spin fixed to a nonmoving substrate. 
Creating strong, well-controlled, local gradients remains challenging for such setups.
In contrast, as we now describe, in CNTs the spin-mechanical coupling
is {\it intrinsic}, supplied by the inherent strong 
spin-orbit coupling \cite{Ando,Jeong,Izumida,Klinovaja}
which was recently discovered by Kuemmeth {\it et  al.}  \cite{kuemmeth08}. 

%
Consider an electron localized in a suspended CNT quantum dot (see Fig.~\ref{fig:setup}).
Below we focus on the case of a single electron, but 
 expect the qualitative features to be valid for any odd occupancy (see Ref. \cite{Jespersen-cntspinorbit}).
%
We work in the experimentally-relevant parameter regime where the spin-orbit and orbital-Zeeman couplings are small compared with the 
nanotube bandgap and the energy of the longitudinal motion in the QD.
Here, the longitudinal and sublattice orbital degrees of freedom are effectively frozen out, leaving behind a nominally four-fold degenerate low-energy subspace associated with the remaining spin and valley degrees of freedom (see Refs. \cite{Palyi-valleyresonance,suppmat}).
\begin{figure}
	\begin{center}
		\includegraphics[width=0.45 \textwidth]{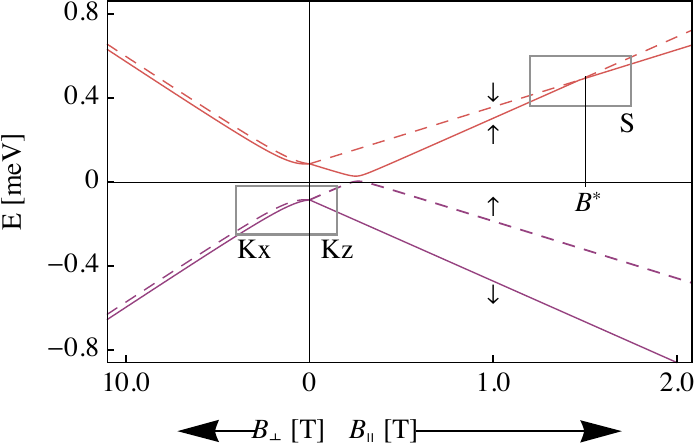}
	\end{center}
	\caption{
	 Energy levels of the four dimensional (due to spin and valley) orbital 
	 ground state subspace of the QD, as a function of 
	 the magnetic field parallel ($B_\parallel$) and
          perpendicular ($B_\perp$) to the CNT
          axis.  The boxed areas indicate the working regime for the
spin qubit (S) and Kramers qubit (K), the latter being operated either
in a longitudinal (Kz) or perpendicular (Kx) magnetic field.
	Parameter values \cite{churchill09}: 
	$\Delta_{\rm so}=170\, \mu {\rm eV}$, 
	$\Delta_{\rm KK'} = 12.5\, \mu {\rm eV}$,
	$\mu_{\rm orb} = 330\, \mu {\rm eV}/{\rm T}$.}
	\label{fig:qubit_spectrum}
\end{figure}

A simple model describing the spin and valley dynamics in this 
low-energy QD subspace, 
incorporating the 
coupling of electron spin to deflections associated with the flexural 
modes of the CNT \cite{Mahan, mariani09}, was introduced in Ref. \cite{Rudner-deflectioncoupling}.
In principle, the deformation-potential spin-phonon coupling mechanism \cite{bulaev08} is also present. 
The deflection coupling mechanism is expected to dominate at long phonon wavelengths, while the deformation-potential coupling should dominate at short wavelengths (see discussion in \cite{Rudner-deflectioncoupling}).
For simplicity we consider only the deflection coupling mechanism, but note that the approach can readily be extended to include both effects.

The Hamiltonian describing this system is \cite{suppmat,Rudner-deflectioncoupling,Flensberg-bentnanotubes}
\begin{equation}
H = \frac{\Delta_{\rm so}}{2} \tau_3 ({\bf s}\cdot {\bf t}) + \Delta_{KK'} \tau_1
- \mu_{\rm orb} \tau_3  ({\bf B}\cdot {\bf t})  + \mu_B (\vec s \cdot \vec B),
\label{eqn:H0}
\end{equation}
where $\Delta_{\rm so}$ and $ \Delta_{KK'} $  denote the spin-orbit 
and intervalley couplings,
$\tau_i$ and $s_i$ are the Pauli matrices in valley
and spin space (the pseudospin is frozen out for the states
localized in a QD),
${\bf t}$ is the tangent vector along the CNT axis, 
and ${\bf B}$ denotes the magnetic field.
Note that the spin-orbit coupling has contributions which are diagonal and off-diagonal in sublattice space 
\cite{Jeong,Izumida,Klinovaja,Jespersen-cntspinorbit}.  
When projected onto to a single longitudinal mode of the quantum dot, the effective Hamiltonian given above describes the coupling of the spin to the nanotube deflection at the location of the dot \cite{suppmat}.

For a nominally straight CNT
we take ${\bf t}$ pointing along the 
$z$ direction, giving ${\bf s}\cdot {\bf t} =s_z$ and ${\bf B}\cdot {\bf
  t} =B_z$.
Here we find the low-energy spectrum shown in Fig.~\ref{fig:qubit_spectrum}.
The two boxed regions indicate two different two-level systems that
can be envisioned as qubit implementations
in this setup: we define a spin qubit \cite{loss-divincenzo} (S)
at strong longitudinal magnetic field, near the value $B^*$ of the upper level crossing, and
a mixed spin-valley or Kramers (K) qubit \cite{Flensberg-bentnanotubes}, 
which can be operated at low fields applied either in the longitudinal (Kz) or perpendicular (Kx) directions. 

We now study how these qubits couple to the quantized mechanical motion of the CNT.
For simplicity we consider only a single polarization of flexural
motion (along the $x$-direction), assuming that the two-fold degeneracy 
is broken, e.g., by an external electric field.
A generalization to two modes is straightforward.

A generic deformation of the CNT with deflection $u(z)$ 
makes the tangent vector ${\bf t}(z)$ coordinate-dependent.
Expanding ${\bf t}(z)$ for small deflections, we rewrite the coupling terms in Hamiltonian (\ref{eqn:H0}) as 
${\bf s}\cdot {\bf t}  \simeq s_z + (du/dz) s_x$ and ${\bf B}\cdot {\bf t}  \simeq B_z + (du/dz) B_x$.
Expressing the deflection $u(z)$ in terms of the creation and annihilation operators $a^\dagger$ and $a$ for a quantized flexural phonon mode, $u(z) = f(z) \frac{\ell_0}{\sqrt 2} (a+a^\dag)$, where
$f(z)$ and $\ell_0$ are the waveform and zero-point amplitude
of the phonon mode, we find that each of the three qubit types
(S, Kx, Kz) obtains a coupling to the oscillator mode which we describe as
\begin{equation}
\frac{H}{\hbar} = 
	\frac{\omega_{q}}{2} \sigma_3 +
	 g  (a+a^\dag) \sigma_1 +
	\omega_p a^\dag a +
	2 \lambda (a+a^\dag) \cos\omega t.
	\label{eq:jcm-1}
\end{equation}
Here the matrices $\sigma_{1,3}$ are Pauli matrices acting on the
two-level qubit subspace, 
and we have included a term describing external driving of the
oscillator with frequency $\omega$ and coupling strength $\lambda$, 
which can be achieved by coupling to the ac electric field 
of a nearby antenna \cite{steele09}.
Below we describe the dependence of the qubit-oscillator coupling $g$ on system parameters for each qubit type (S, Kx, or Kz).
The derivation of Eq.~(\ref{eq:jcm-1}) is detailed in \cite{suppmat}.

For the spin qubit (S), the relevant two-fold degree of freedom is the spin of the electron itself.
Therefore in Eq.(\ref{eq:jcm-1}) we have $\sigma_3=s_z$ and $\sigma_1=s_x$, and the qubit
levels are split by the Zeeman energy, measured relative to the value $B^*$
where the spin-orbit-split levels cross, $\hbar\omega_q = \mu_B(B-B^*)$.  
A spin magnetic moment of $\mu_B$ is assumed, and $B^* \approx \Delta_{\rm so}/2\mu_B$
for $\Delta_{KK'} \ll \Delta_{\rm so}$. 
For the qubit-resonator
coupling, we find $g=\Delta_{\rm so}\langle f'\rangle\ell_0/2\sqrt{2}$, 
independent of $B$. 
Here, $\langle f'\rangle$ is the derivative of the waveform of 
the phonon mode averaged against the electron density profile in the QD.

For a symmetric QD, positioned at the midpoint of the CNT, the coupling matrix element proportional to $\langle f'\rangle$ vanishes for the fundamental and all even harmonics (the opposite would be true for the deformation-potential coupling mechanism).
The cancellation is avoided 
for a QD positioned away from the symmetry point of the CNT, or 
for coupling to odd harmonics.
Here, for concreteness, we consider coupling of a symmetric QD to the first vibrational harmonic of the CNT.   
Using realistic parameter values 
\cite{kuemmeth08,churchill09,Huttel-cntresonator, steele09},
$L = 400$ nm, $\ell_0 = 2.5$ pm, $\Delta_{\rm so} = 370\ \mu$eV,
$\Delta_{KK'} = 32.5 \ \mu$eV, $\mu_{\rm orb} = 1550 \ \mu{\rm eV}/{\rm T}$, and
$\omega_p/2\pi = 500$ MHz,  
we find $g/2\pi \approx 0.56$ MHz, irrespective of the magnetic field strength $B$ along the CNT. 
\begin{figure*}[ht]
	\includegraphics[width=7.1 in]{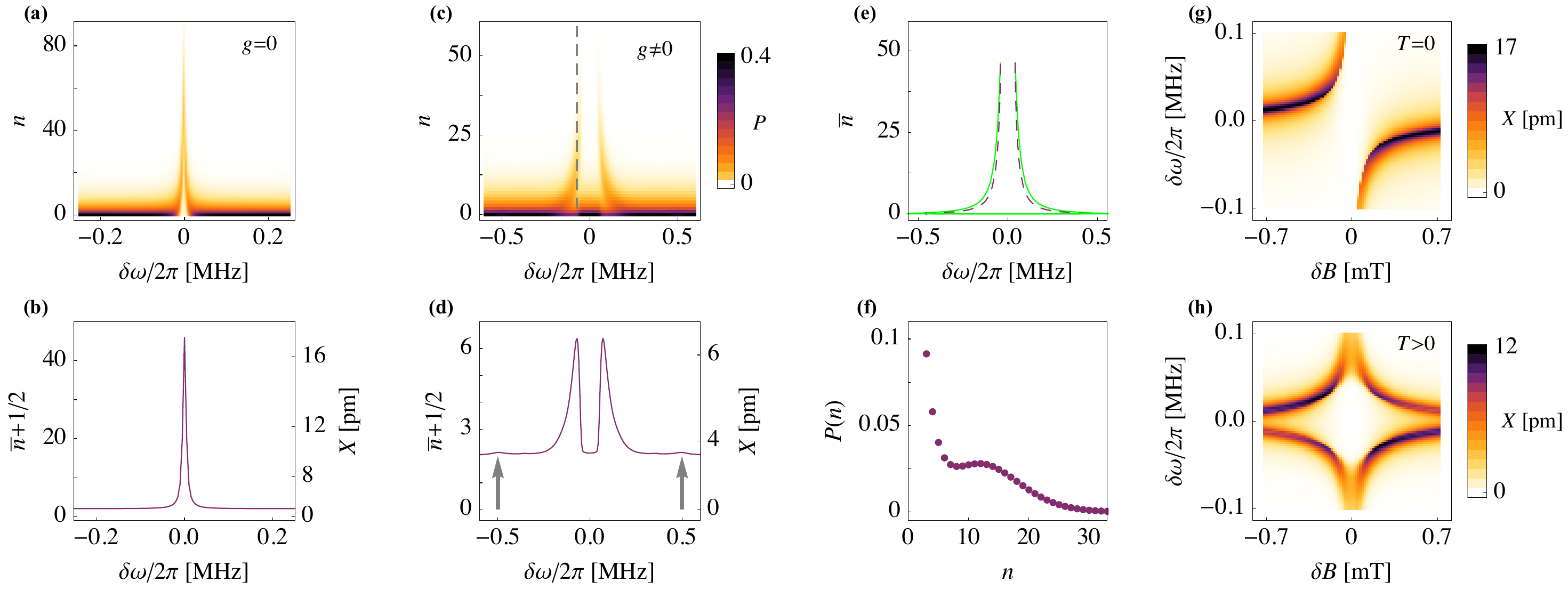}
	\caption{Response of the spin-oscillator system.
        (a) Phonon number probability distribution $P(n,\delta\omega)$,
	(b) average phonon occupation $\bar n$
	and root mean squared displacement $X$ 
	of the uncoupled driven CNT resonator ($g=0$), 
        as functions of the drive frequency--oscillator frequency
	detuning $\delta \omega = \omega-\omega_p$.
	The parameters are $T=50$ mK, $\omega_p/2\pi = 500\ {\rm MHz}$,
	$\Gamma = 5\cdot 10^4\ {\rm s}^{-1}$ and
	$\lambda/2\pi = 0.027\ {\rm MHz}$.
	The same quantities are plotted in (c) and (d) for
	a resonantly coupled qubit-oscillator system (i.e., $\omega_q = \omega_p$),
	with coupling constant 
        $g/2\pi = 0.5\ {\rm MHz}$ and
        further parameters as in (a) and (b).  
	(e) Steady-state oscillator response from the semiclassical calculation,
	corresponding to the parameters of (c) and (d).
        The green solid (purple
        dashed) lines describe stable (unstable) solutions.  
	(f) Bimodal phonon number distribution, taken along the dashed vertical
	line of (c). 
	(g,h) Root mean squared value $X$ of the resonator amplitude 
        in the coupled spin qubit - oscillator system at
	(g) $T=0$ and (h) $T=50$ mK,
	as functions of magnetic field detuning $\delta B$ (detuning
        the qubit frequency away from resonance with the oscillator)
        and drive frequency--oscillator frequency detuning $\delta \omega$.
	}
	\label{fig:results}
\end{figure*}

For the Kramers qubits (Kx and Kz), both $\omega_q$ and $g$ depend on $B$. 
The qubit splitting for the Kx qubit is controlled by
the perpendicular field, $\hbar\omega_q = \mu_B (2\Delta_{KK'}/\Delta) B_x$,
while for the Kz qubit, it is controlled by the longitudinal field $\hbar\omega_q = (\mu_B+\mu_{\rm orb} (\Delta_{\rm so}/\Delta)) B_z$,
where $\Delta=\sqrt{\Delta_{\rm so}^2 + 4\Delta_{KK'}^2}$ denotes
the zero-field splitting between the two Kramers pairs.
Resonant coupling occurs when $\omega_q = \omega_p$. 
This condition sets the relevant value of $B_x$ ($B_z$) 
in the case of the Kx (Kz) qubit; the parameters above yield 
$B_x \approx 103\ {\rm mT}$ ($B_z \approx 0.6\ {\rm mT}$).

The qubit-cavity coupling for the Kx qubit
increases linearly with the applied perpendicular field,
$\hbar g= - (\langle f'\rangle\ell_0/\sqrt{2}) (\mu_{\rm
  orb}\Delta_{\rm so}/\Delta +\mu_B \Delta_{\rm so}^2/\Delta^2 )
B_x$,
while for the  Kz qubit it scales with the longitudinal field,
$\hbar g=  (\langle f'\rangle\ell_0/\sqrt{2}) 
(\mu_{\rm orb}2\Delta_{KK'}\Delta_{\rm so}/ \Delta^2 ) B_z$.
Using the values of 
$B_x$ and $B_z$ 
obtained above, we estimate couplings of $g/2\pi \approx 0.49\ {\rm MHz}$ for the Kx qubit, and $g/2\pi \approx 0.52\ {\rm kHz}$ for the Kz qubit.
Thus the coupling for the Kx qubit is comparable to that of the spin qubit, while the coupling of the Kz qubit is much weaker.
Therefore, we restrict our considerations to the spin and Kx qubits below. 

Ref.~\onlinecite{steele09} reports the fabrication of CNT
resonators with quality factors $Q \approx 150\mathord{,}000$. 
We take $Q = 63\mathord{,}000$ for the following estimate.
Together with the oscillator frequency $\omega_p/2\pi = 500\ {\rm MHz}$, 
this value of $Q$ implies an oscillator damping rate of 
$\Gamma \approx 5\cdot 10^4\ {\rm s}^{-1}  \ll g$.
Because of the 
near-zero density of states of other phonon modes at $\omega_q$, 
it is reasonable to assume a very low spontaneous qubit relaxation rate $\gamma$.
These observations suggest that the so-called ``strong coupling'' regime of
qubit-oscillator interaction, defined as $\Gamma,\gamma \ll g$, can be reached with CNT resonators. 

To quantify the system's response in the anticipated parameter regime, we study the coupled qubit-oscillator dynamics using a master equation which takes into account the finite lifetime of the phonon mode as well as the non-zero temperature of the external phonon bath.
For weak driving, $\lambda \ll \omega_p$, and
$\omega_p \approx \omega_q \approx \omega \gg g$, 
we move to a rotating frame and use the rotating wave approximation (RWA) to 
%
map the Hamiltonian, Eq.(\ref{eq:jcm-1}), into Jaynes-Cummings form \cite{Jaynes}
\begin{equation}
	\frac{H_{\rm RWA}}{\hbar} = 
	\frac {\tilde\omega_{q}}{2}  \sigma_3 +
	g  (a\sigma_+ + a^\dag \sigma_-) +
	\tilde\omega_p a^\dag a +
	\lambda (a+a^\dag),
\label{eq:jcm}
\end{equation}
where $\tilde\omega_i=\omega_i-\omega$.
Including the nonunitary dynamics associated with the phonon-bath coupling, the master equation for the qubit-oscillator density matrix $\rho$ reads: 
\begin{eqnarray}
\dot \rho =  - \frac i \hbar \left[ H_{\rm RWA} , \rho\right] 
     & +&  (n_B + 1) \Gamma\,
	\big( a\rho\adag - \frac{1}{2}\{\adag a,\rho\}\big) \nonumber\\
& + &
	n_B \Gamma\,
	\big( \adag\rho a - \frac{1}{2}\{a\adag,\rho\}\big),
	\label{eqn:lindblad_terms}
\end{eqnarray}
where $n_B = 1/(e^{\hbar\omega_p/k_B T}-1)$ is the bath-mode Bose-Einstein occupation factor, 
and $k_B$ is the Boltzmann constant. 

Because of the phonon damping, in the long-time limit the system is expected to tend towards a steady state, described by the density matrix $\bar \rho$.
We study these steady states, found by setting $\dot{\rho} = 0$ in
Eq.(\ref{eqn:lindblad_terms}), using both numerical and semiclassical analytical methods. 
In Figs.~\ref{fig:results}a,c  we show the 
steady-state phonon occupation probability distribution $P(\delta \omega,n)$ 
as a function of the drive frequency--phonon frequency detuning 
$\delta \omega = - \tilde \omega_p$ 
and the phonon occupation number $n$, 
for the case where the qubit and oscillator frequencies are fixed and 
degenerate, $\omega_q = \omega_p$ (see caption for parameter values).
Panels a and c compare the cases with and without qubit-oscillator coupling.
In Figs.~\ref{fig:results}b and \ref{fig:results}d we show the averaged phonon occupation number $\bar{n}(\delta \omega) = \sum_n nP(\delta \omega, n)$, which is  
closely related to the mean squared resonator displacement 
in the steady state: $X ^2 = \overline{x^2} = \ell_0^2(\bar n + \frac 1 2)$.
For $g\neq 0$, we observe 
a splitting of the oscillator resonance, which is characteristic of the coupling to the 
two-level system, and can serve as an experimental signature
of the qubit-oscillator coupling. 
%
%
%
For drive frequencies near the split peaks, the phonon number distribution
is bimodal (Fig.~\ref{fig:results}f) showing peaks at $n\approx 0$ and at high-$n$, 
indicating bistable behavior (see below).

For strong excitation, where the mean phonon occupation is large, we expect a semiclassical approach to capture the main features of the system's dynamics \cite{AlsingCarmichael,Girvin}.
Extending the approach described in \cite{AlsingCarmichael} to include distinct values of the qubit, oscillator, and drive frequencies, $\omega_q, \omega_p$, and $\omega$, we derive semiclassical equations of motion for the mean spin and oscillator variables (see \cite{suppmat}).
The steady-state values of the mean squared oscillator amplitude obtained 
from the resulting nonlinear system are shown in Fig.~\ref{fig:results}e.
In the vicinity of the split peak we find two branches of stable steady-state solutions, 
indicative of bistable/hysteretic behavior \cite{steele09}.
The semiclassical results in Fig.~\ref{fig:results}e are in correspondence with the
phonon number distribution in Fig.~\ref{fig:results}c,
and explain its bimodal character.
Similar oscillator instabilities have been used as the basis for a sensitive readout scheme in superconducting qubits \cite{Reed}, and may potentially be useful for mass or magnetic field sensing applications where small changes of frequency need to be detected.  

To predict the oscillator response to be detected via a charge
sensor (see below), we solve for the stationary state of 
Eq.~(\ref{eqn:lindblad_terms}) directly
for a range of driving frequencies, qubit-oscillator detunings (set by the
magnetic field), and temperatures $T$.
In Figs.~\ref{fig:results}(g) and \ref{fig:results}(h), 
we show the $T=0$ and $T = 50$ mK
root mean squared oscillator amplitude $X \propto \sqrt{{\bar n}+1/2} $ 
as function of magnetic field $B$ and drive frequency, for the case of a 
spin (S) qubit.
The value $\delta B =0$ corresponds to resonant coupling $\omega_q = \omega_p$.
These results also apply for the Kx qubit, if the magnetic field axis is adjusted appropriately. 
In the zero-temperature case, only half of the eigenstates 
$\hbar \omega_{\pm} \approx  
\hbar \omega_p \mp \hbar{g^2}/(\omega_p-\omega_q)$
of Eq.~(\ref{eq:jcm}) can be efficiently excited by the drive at fixed
$\delta B$, giving
rise to the upper (lower) feature in Fig.~\ref{fig:results}g for
$\delta B<0$ ($\delta B>0$).
However, for $T \gtrsim \hbar\omega_q$, both branches of the Jaynes-Cummings
ladder can be efficiently excited (Fig.~\ref{fig:results}h).  This is
a distinct and experimentally accessible signature of the strong
coupling at finite temperature.
Note that the 
vacuum Rabi splitting 
is also observed (see arrows in Fig.~\ref{fig:results}d), but features arising from nonlinearity in the strongly 
driven system dominate by more than 2 orders of magnitude.

%
Displacement detection of nanomechanical systems is possible using charge
sensing \cite{Chiu08, Knobel-nems-set},
where the conductance of a mesoscopic conductor, such as a QD or quantum point contact, is modulated via capacitve coupling to the charged mechanical resonator. 
Furthermore, 
the qubit state itself can be read out using spin-detection schemes developed for semiconductor QDs \cite{Elzerman04}, or by a dispersive readout scheme like that commonly used in superconducting qubits coupled to microwave resonators \cite{Wallraff05}.
The dispersive regime can be rapidly accessed by, e.g., tuning the resonator frequency using dc gate pulses which control the tension in the CNT \cite{steele09}.

%
In summary, 
we predict that strong qubit-resonator coupling 
can be realized in suspended CNT QDs with current state-of-the-art devices. 
The coupling
described here may find use in sensing applications, and in spin-based quantum information processing, where the CNT oscillator enables electrical control of the electron spin, and, with 
capacitive couplers, may provide long-range interactions between distant electronic qubits \cite{Imamoglu,Rabl-spinphonon}. 
Combined with control of the qubit via electron-spin-resonance \cite{Koppens}, the mechanism studied here could be utilized for ground-state cooling and for generating arbitrary motional quantum states of the oscillator \cite{Rabl-strongcoupling}.

We gratefully acknowledge helpful discussions with H. Carmichael, V. Manucharyan and P. Rabl.
This work was supported by the OTKA grant PD 100373, the
Marie Curie grant CIG-293834 and the QSpiCE program of ESF (AP),
DFG under the programs FOR 912 and SFB 767 (PS and GB),
NSF grants DMR-090647 and PHY-0646094 (MR), 
and The Danish Council for Independent Research | Natural Sciences (KF).

\textit{Note:} While completing this manuscript, we became aware of a
related work \cite{wegewijs} that describes the theory of the spin-phonon coupling 
in a CNT resonator QD, and its consequences in the spin blockade transport setup.


\newpage
\setcounter{figure}{0}
\setcounter{equation}{0}
\setcounter{section}{0}


%


\onecolumngrid

\begin{center}
\large{{\it Supplementary electronic material for:}\\
Spin-Orbit-Induced Strong Coupling of a Single Spin to
a Nanomechanical Resonator}
\end{center}

\begin{center}
Andr\'as P\'alyi,$^{1,2}$
P.~R.~Struck,$^1$
Mark Rudner,$^3$
Karsten Flensberg,$^{3,4}$
and
Guido Burkard$^1$
\end{center}

\begin{center}
{\it 
$^\mathit{1}$Department of Physics, University of Konstanz, D-78457 Konstanz, Germany\\
$^2$Department of Materials Physics, E\"otv\"os University,
H-1517 Budapest POB 32, Hungary\\
$^3$Department of Physics, Harvard University, Cambridge, Massachusetts 02138, USA\\
$^4$Niels Bohr Institute, University of Copenhagen, Universitetsparken 5, DK-2100 Copenhagen, Denmark
}
\end{center}

%
%
%
%
%

\section{Appendix A: Effective quantum dot Hamiltonian}

In the following, we derive Hamiltonian (1) of the main text, which describes the
spectrum of a quantum dot formed in a straight, suspended carbon nanotube, 
in the absence of phonons (i.e.~in a static tube). 
As usual, we start from
a tight-binding description of a graphene sheet, which is then rolled up
with the condition of periodic boundary conditions. Using the
conventions as in Weiss {\em et al.} \cite{Weiss}, this gives the following Hamiltonian for the longitudinal degree of freedom
\begin{equation}\label{H}
	H_{0}= v_{\mathrm{F}}p_{z}\sigma_{2}+\Delta_{g}^{{}}\sigma_{1}^{{}}+\hat{\mathbf{t}}\cdot \mathbf{s}\,
\tau_{3}^{{}}(\sigma_{1}^{{}}\Delta_{1}^{{}}+\Delta_{0}^{{}})+V(z).
\end{equation}
Here $z$ and $\hat{\mathbf{t}}$  represent the coordinate and unit vector in the direction of the tube, $v_F$ is the Fermi velocity, $\sigma_i$,$\tau_i$, and $s_i$ are Pauli matrices in sublattice, valley and spin spaces, respectively. 
Note that to translate between the convention used here and that used in e.g. Ref.~\onlinecite{Klinovaja2011}, $\sigma_1$ must be replaced by $\tau_3\sigma_1$.
The energy gap between the valence band and the conduction band is
$2\Delta_g$, where
$\Delta_g=\hbar v_F(\nu/3R)+\Delta_c$ with $2\Delta_c$ being the curvature induced minigap, which is typically of order 10 meV, but proportional to $\cos3\theta$, where $\theta$ is the chiral angle of the tube. For nominally metallic tubes, $\nu=0$. The spin-orbit interaction has two terms, one that is diagonal in sublattice (the $\Delta_0$ term) and one which is off-diagonal (the $\Delta_1$ term). The spin-orbit interaction connects the spin projection along the tube axis to the $K,K'$ (valley flavor) quantum number through the prefactor $(\hat{\mathbf{t}}\cdot\mathbf{s})\tau_{3}$. The microscopic derivation of this Hamiltonian can be found in Refs.~\onlinecite{Jeong2009,Izumida2009,Klinovaja2011}. Finally, the term $V(z)$ describes the confining potential of the quantum dot. It is assumed to be smooth on the atomic scale and hence has no sublattice structure.

Furthermore, an applied magnetic field couples to both the spin and orbital degrees of freedom. The coupling to the orbital degree of freedom appears through the Aharonov-Bohm flux, which modifies the boundary condition of the circumferential wave vector and hence changes $\Delta_g$. In total, the Hamiltonian due to magnetic field is
\begin{equation}
H_{B}=H_{orb}+H_{s},\quad H_{orb}=-\mu_{B}(l/\hbar)\sigma_1\tau_{3}\mathbf{\hat{t}~}\cdot\mathbf{B},\quad
H_{s}=\frac{1}{2}g\mu_{B}~\mathbf{s}\cdot\mathbf{B,}%
\end{equation}
where $l=m v_{\mathrm{F}}R$.

There are several ways to arrive at the Hamiltonian in Eq.~(1) in the main text. Here we will assume a hierarchy of energy scales, typical of many experimental realizations of nanotube quantum dots, namely
\begin{equation}\label{scales}
    \Delta_g \gg E_L \gg  E_B, E_\mathrm{SO},E_{KK'}, 
\end{equation}
where $E_L$ is the level spacing due to longitudinal quantization and $E_B$, $E_\mathrm{SO}$, and $E_{KK'}$ are the energy changes due to the external magnetic field, spin-orbit coupling, and valley mixing, respectively. This allows us to first solve for the dot wavefunction in absence of these three contributions and then project onto a single longitudinal mode. 

In passing we note that in order to get more information about
dependence of the orbital magnetic moment and spin-orbit coupling on the 
number of electrons in the quantum dot, one has to be more precise and use a specific form of the confining potential, e.g. assuming a square well potential, as in Refs.~\onlinecite{Bulaev,Weiss}. This was done in Refs.~\onlinecite{Jespersen1,Jespersen2}, where a method to experimentally extract the two spin-orbit parameters $\Delta_0$ and $\Delta_1$, as well as $\mu_\mathrm{orb}$, was shown.

Now imagine that one has solved for the case without magnetic field, spin-orbit coupling, and mixing between $K$ and $K^{\prime}$. This gives a set of longitudinal wavefunctions, each one four-fold degenerate due to the spin and valley degrees of freedom. The energy splitting between these shells is $E_L\gg E_B, E_\mathrm{SO},E_{KK'}$.
We label these states by the valley and spin quantum numbers $\tau = \pm 1$ and $s = \pm 1$, respectively, which indicate corresponding eigenvalues under $\tau_3$ and $s_z$.
Projected onto eigenstates of the spatial coordinates $z$ and $c$ (the circumferential coordinate), the wave function in the
envelope-function representation  
has the form\cite{Weiss} 
\begin{equation}
\langle (z,c)|\tau,s\rangle =\frac{e^{i\tau k_\perp c}}{\sqrt{2\pi R}}
\phi(z)\eta_\sigma\otimes\chi_s
\otimes \chi_\tau,
\label{psi}
\end{equation}
where $k_\perp = -\frac \nu {3R}$ is the wave vector associated with the  
gap, $\eta_\sigma = \frac{1}{\sqrt 2}(1\ 1)^{\rm T}$ is a pseudospinor describing the sublattice degrees of freedom, 
$\chi_s$ and $\chi_\tau$ describe the spin and valley degrees of freedom,
respectively,
$\chi_+ = (1\ 0)^{\rm T}$ and 
$\chi_- = (0\ 1)^{\rm T}$,
and the precise form of the envelope wave function $\phi(z)$
depends on the confining potential. 

We can now take matrix elements with respect to $H_B$ and $H_\mathrm{SO}=\mathbf{\hat{t}\cdot s}
\tau_{3}^{{}}(\sigma_{1}^{{}}\Delta_{1}^{{}}+\Delta_0)$ and also a term describing the microscopic disorder that couples valleys: $H_{KK'}=V_{KK'}$, where $V_{KK'}$ is a short-range disorder potential depending on the longitudinal and circumferential coordinate operators. 
(Note that short-range disorder can be systematically incorporated into
the envelope-function description, see e.g., 
Refs. \onlinecite{Ando-impurity,Palyi-valleyresonance}.)
This procedure will produce a Hamiltonian of the form in Eq.~(1) in the main text,
\begin{equation*}
H = \frac{\Delta_{\rm so}}{2} \tau_3 ({\bf s}\cdot {\bf t}) + \Delta_{KK'} \tau_1
- \mu_{\rm orb} \tau_3  ({\bf B}\cdot {\bf t})  + \mu_B (\vec s \cdot \vec B),
\end{equation*}
 with
\begin{eqnarray}
 \Delta_\mathrm{so} = 2 \langle +|\sigma_{1}^{{}}\Delta_{1}^{{}}+\Delta_0|+\rangle, \ \ 
  \Delta_{KK'} =  \langle -|V_{KK'}|+\rangle, \ \
    \mu_\mathrm{orb} =  \frac{e v_{\mathrm{F}}R}{2}\langle +|\sigma_{1}|+\rangle.
\end{eqnarray}
Here the kets $|\pm\rangle$ stand for the orbital states with $\tau = \pm 1$.
Note that in general, the valley-mixing term can include both $\tau_1$ and
$\tau_2$, but an appropriate unitary transformation in the valley space can be used to put it into 
the form above with real $\Delta_{KK'}$.

\subsection{A.1 Derivation of the coupling to vibrations}

Next we look at how the vibrations couple to the four states of the quantum dot. The amplitude of the tube is in terms of the harmonic oscillator raising/lowering operators given by
\begin{equation}
u(z)=f(z)\frac{\ell_0}{\sqrt{2}}\left(  a+a^{\dagger}\right)  ,
\end{equation}
where we focus on a single vibrational mode. The coupling to the
spin is via the change of the tangent direction given by
\begin{equation}\label{dz}
\delta\mathbf{\hat{t}=}\frac{du}{dz}\mathbf{\hat{x},}%
\end{equation}
where $\mathbf{\hat{x}}$ is perpendicular to the tube and in the plane of the
vibration. The interaction Hamiltonian then becomes%
\begin{equation}
H_{s,\mathrm{vib}}=\delta\mathbf{\hat{t}\cdot s~}\tau_{3}^{{}}(\sigma_{1}^{{}%
}\Delta_{1}^{{}}+\Delta_{0}^{{}})-\mu_{\mathrm{orb}}\tau_{3}^{{}}\sigma_{1}\delta\mathbf{\hat{t}\cdot B.}%
\end{equation}
As above, we now project onto a single longitudinal mode, thus taking matrix elements of $H_{s,\mathrm{vib}}$ in the basis $|\tau,s\rangle$. Such matrix elements involve form factors like
\begin{equation}
\langle \tau,s|\sigma_{i}^{{}}f^{\prime}(z)|\tau',s'\rangle=\delta_{\tau\tau^{\prime}}\langle \tau,s|\sigma_{i}^{{}}f^{\prime}(z)|\tau,s'\rangle=\delta_{\tau\tau^{\prime}} \delta_{ss'} F_{i,\tau}.
\end{equation}
At this point we note that the coupling is small for 
a symmetric dots and even harmonics, because the $F$ factors then tend to cancel, see discussion in the main text. The effective Hamiltonian for coupling between the four states of the quantum dot and the vibration now becomes
\begin{equation}\label{Hsvib}
H_{s,\mathrm{vib}}=\frac{\ell_0}{\sqrt{2}}\left(  a+a^{\dagger}\right)  \left\{
s_x \tau_{3}^{{}}(F_{1}\Delta_{1}^{{}}+F_{0}\Delta_{0}^{{}})-B_x\tau_{3}^{{}}\mu_{\mathrm{orb}}F_{1}\right\}.
\end{equation}

We see from Eq.~\eqref{Hsvib} that the coupling of the vibrations to quantum dot states have different form factors from what one would get by simply setting \eqref{dz} into Eq.~(1) of the main paper. However, when the energy scales are clearly separated as in \eqref{scales} the eigenstates $|\tau,s\rangle$ are eigenstates of $\sigma_1$ (which can be seen from \eqref{H}) and therefore $F_0=F_1$ (for the conduction band). Therefore, we do not need to take into account the different form factors in \eqref{Hsvib}, which simplifies the analysis  and we can write $F_0=F_1=\langle f' \rangle$. In this language Eq.~\eqref{Hsvib} becomes
\begin{equation}\label{Hsvib2}
    H_{s,\mathrm{vib}}=\frac{\ell_0}{\sqrt{2}}\left(  a+a^{\dagger}\right)  \left\{
s_x\tau_{3}^{{}}\Delta_\mathrm{SO}-B_x \tau_{3}^{{}}\mu_{\mathrm{orb}}\right\}\left\langle f' \right\rangle,
\end{equation}
which is the result used in the main text.

As mentioned the expression \eqref{Hsvib2} was derived under the assumption that the gap dominates over the longitudinal size quantization energy, which is valid for few-electron quantum dot. However, it is important to note that one could easily extend this to the more general case at higher energies by including the difference in form factors $F_0$ and $F_1$ without changing the conclusions and structure of our results qualitatively.

\section{Appendix B: Qubit-phonon couplings}

We treat three different qubit realizations in the main text: the spin  qubit (S),
the Kramers qubit in a magnetic field perpendicular  to the
carbon nanotube (CNT) (Kx),
and
the Kramers qubit in a parallel-to-CNT magnetic field (Kz).
Below, we express the three qubit Hamiltonians
$\bar H_s$, $\bar H_{Kx}$ and $\bar H_{Kz}$
as functions of system parameters and CNT deformation.
To clarify the correspondence with the Jaynes-Cummings
Hamiltonian in Eq. (2) of the main text, we
list the formulas for the qubit frequency, the qubit-phonon coupling,
as well as numerical estimates for the latter, in
Table \ref{tab:jcm-correspondence}.

\subsection{B.1 Spin-phonon coupling}
At the finite value
\begin{equation}
B^* = \frac{\Delta_{\rm so}}{2\mu_B}
\sqrt{1-\frac{4\Delta_{KK'}^2}{\Delta_{\rm so}^2\left(
\frac{\mu_{\rm orb}^2}{\mu_B^2} - 1
\right)}}
\end{equation}
of a longitudinally-applied magnetic field,
the quantum dot (QD) energy spectrum shows
a crossing of the energies of a pair of spin states belonging to the same valley
(see Fig. 2 of the main text).
Around this point these two levels are energetically well separated from the other
QD levels.
We call this two-level system the spin qubit (S).
If the dynamics is restricted to these two levels, it can
be described by the following effective Hamiltonian:
\begin{equation}
H_s = \mu_B s_z (B-B^*) + \frac{\Delta_{\rm so}} {2}
\hmprime s_x
\end{equation}
where we assume that the effect of valley-mixing is negligible,
$\Delta_{KK'} = 0 $.
Averaging over the $z$ coordinate using the charge density $n(z)$ of
the electron occupying the CNT QD yields
\begin{equation}
\bar H_s \equiv \int dz n(z) H_s (z) = \mu_B s_z (B-B^*) + \frac{\Delta_{\rm so}} {2} s_x
\int dz n(z) \hmprime \equiv
\mu_B s_z (B-B^*) + \frac{\Delta_{\rm so}} {2} s_x
\avg{\hmprime}.
\end{equation}

\subsection{B.2 Kramers qubit-phonon coupling in a perpendicular magnetic field}

At zero magnetic field, the ground state of the CNT QD,
i.e., the ground state of the Hamiltonian $H$ in Eq. (1) of the main text,
is formed by a pair of time-reversed states (Kramers pair).
The twofold degeneracy is maintained even in the
presence of spin-orbit interaction and valley mixing.
At small enough magnetic field these two states split up,
but they remain energetically well separated from higher-lying states.
We call this two-level system the `Kramers qubit' \cite{Flensberg-bentnanotubes}
in a perpendicular field (Kx).
[Similar considerations hold for the first excited Kramers pair, i.e., the
two higher-lying energy eigenstates of $H$ in Eq. (1) of the main text.]
These energetically split states,
in the absence of CNT deformation, will be
denoted here as $\ket{+}$ and $\ket{-}$.
Starting from the Hamiltonian $H$ in Eq. (1) of the main text, averaging
over $z$ using the electron density $n(z)$, and incorporating the effect
of the two higher-lying states on $\ket{+}$ and $\ket{-}$ via
a second-order Schrieffer-Wolff transformation, we find that
the dynamics restricted to the Kramers qubit in the presence of an external
magnetic field $\vec B = B_x \hat{\vec x}$
and CNT deformation $u(z)$ is described by the Hamiltonian
\begin{equation}
\bar H_{Kx} = B_x \left[
	\sigma_3 	\frac{2\mu_B \Delta_{KK'}}{\Enull} -
	\sigma_1 \avg{ \hmprime} \left(
		\frac{\mu_{\rm orb} \Delta_{\rm so}}{\Enull} +
		\frac{\mu_B \Delta_{\rm so}^2}{\Delta_{\rm so}^2+4\Delta_{KK'}^2}
	\right)
\right].
\end{equation}
Here, $\sigma_{1,3}$ are the Pauli matrices in the
qubit basis, i.e.,
 $\sigma_3 = \ket{+} \bra{+} - \ket{-} \bra{-}$ and
 $\sigma_1 = \ket{+} \bra{-} + \ket{-} \bra{+} $.

\begin{table*}
\centering
\begin{tabular}{|l||c|c|c|}
		\hline
			Hamiltonian &
			$\hbar \omega_q$ &
			$\hbar g$ &
                        $g/2\pi$ (numerical)\\
		\hline
			$\bar H_s$ &
			$\mu_B(B-B^*)$ &
			$\frac{\Delta_{\rm so} \avg{f'}
                          \ell_0}{2\sqrt{2} } $ &
                        $0.56\,{\rm MHz}$\\
			$\bar H_{Kx}$ &
			$\frac{\mu_B B_x 2\Delta_{KK'}}{\Enull}$ &
			$-\frac{B_x \avg{f'} \ell_0}{\sqrt 2}
                        \left(\frac{\mu_{\rm orb} \Delta_{\rm so}}{\Enull}
                          +	\frac{\mu_B \Delta_{\rm
                              so}^2}{\Delta_{\rm
                              so}^2+4\Delta_{KK'}^2}\right)$ &
                        $0.49\,{\rm MHz}$\\
			$\bar H_{Kz}$ &
			$- B_z \left(\mu_B + 	\frac{\mu_{\rm orb} \Delta_{\rm so}}{\Enull}\right)$ &
			$\frac{B_z \avg{ f'} \ell_0}{\sqrt
                          2}\frac{\mu_{\rm orb} 2\Delta_{KK'} \Delta_{\rm
                            so}} {\Delta_{\rm so}^2+4\Delta_{KK'}^2}$  &
                        $0.52\,{\rm kHz}$\\
		\hline
\end{tabular}
	\caption{Correspondence between the terms of the
	Hamiltonian in Eq.~(2) of the main text and those of
	the qubit-phonon Hamiltonians derived for the three different
        qubits.
      The parameters used to calculate the last column are
$L = 400$ nm, $\ell_0 = 2.5$ pm, $\Delta_{\rm so} = 370\ \mu$eV,
$\Delta_{KK'} = 32.5 \ \mu$eV, $\mu_{\rm orb} = 1550 \ \mu{\rm eV}/{\rm T}$.
	The estimate $\avg{f'} = 2\sqrt 2 / L$ has been used (see text).}
	\label{tab:jcm-correspondence}
\end{table*}

\subsection{B.3 Kramers qubit-phonon coupling in longitudinal magnetic field}

In the absence of CNT deformation, a parallel-to-CNT
magnetic field splits two low-energy Kramers doublet of $H$ in
Eq. (1) of the main text.
We call this two-level system the Kz qubit, and denote the two qubit states
as $\ket{+}$ and $\ket{-}$ in this subsection.
Starting from the complete Hamiltonian $H$ in Eq. (1) of the main text, averaging
over $z$ using the electron density $n(z)$, and applying
a second-order Schrieffer-Wolff transformation to describe
the effect of the higher-lying Kramers pair to the Kz qubit, we find that
the dynamics of the latter in the presence of an
external magnetic field $\vec B = B_z \hat{\vec z}$ and
CNT deformation $u(z)$ is described by the Hamiltonian
\begin{equation}
\bar H_{Kz} = B_z \left[
	\sigma_1 \avg{ \hmprime}
		\frac{\mu_{\rm orb} 2\Delta_{KK'} \Delta_{\rm so}}
		{\Delta_{\rm so}^2+4\Delta_{KK'}^2} -
	\sigma_3  \left(
		\mu_B +
		\frac{\mu_{\rm orb} \Delta_{\rm so}}{\Enull}
	\right)
\right].
\end{equation}
As before, $\sigma_{1,3}$ are the Pauli matrices in the
 qubit basis, i.e.,
 $\sigma_3 = \ket{+} \bra{+} - \ket{-} \bra{-}$ and
 $\sigma_1 = \ket{+} \bra{-} + \ket{-} \bra{+} $.

\subsection{B.4 Deformation of the CNT}

All three qubit-phonon Hamiltonians $\bar H_s$, $\bar H_{Kx}$ and
$\bar H_{Kz}$ resemble the Jaynes-Cummings Hamiltonian of
cavity quantum electrodynamics.
This becomes more apparent if we express the $z$-dependent displacement
$u(z)$ in
terms of phonon annihilation $a$ and creation $a^\dag$ operators:
\begin{equation}
u(z) = f(z) \frac{\ell_0}{\sqrt 2} (a+a^\dag).
\end{equation}
Here $f(z)$ is a dimensionless function describing the shape of the
standing-wave bending phonon mode
under consideration (normalization: $\int f^2(z) dz = L$)
and $\ell_0$ is the ground-state displacement of that mode.

As apparent from Table \ref{tab:jcm-correspondence}, the qubit-phonon
coupling vanishes if
$g \propto \avg{f'} \equiv \int dz \frac{df(z)}{dz} n(z) = 0$.
Therefore, $g$ vanishes if the setup is perfectly left-right symmetric
along the CNT ($z$) axis and a bending mode with an even number of
nodes is considered.
To have a finite qubit-phonon coupling, either the left-right symmetry
of
the setup must be broken or a flexural mode with odd number of nodes
should be considered.
In the main text and also here
we treat the second case: we investigate the coupling of the
first excited flexural phonon (1 node) to the various qubits.

The displacement field of the first harmonic of the resonator can be approximated by
\begin{equation}
f(z) = - \sqrt{2} \sin \left[ \frac{2\pi} L \left(z+\frac L 2\right) \right],
\end{equation}
where we assume that the CNT is suspended at points $z=-L/2$ and
$z=L/2$.
Approximating the charge density with a step function symmetrically covering
the length fraction $\xi$ of the suspended part of the CNT, we obtain
\begin{equation}
\avg{f'} = - \int_{-\xi L /2}^{\xi L/2} dz \frac 1 {\xi L}  \frac{2\pi}{L} \sqrt 2
\cos \left[ \frac{2\pi} L \left(z+\frac L 2\right) \right] =
\frac{2 \sqrt 2} L \frac{\sin \pi \xi} \xi
\end{equation}
The fraction $\sin (\pi \xi) / \xi$ is $\sim 1$ if $\xi$ is smaller than 1, and therefore
we make the approximation $\avg{f'} \approx 2\sqrt{2}/L$ in the numerical
estimates appearing in the main text and in Table \ref{tab:jcm-correspondence}.

\section{Appendix C: Semiclassical equations of motion}
In this section we develop semiclassical equations of motion for the coupled qubit-oscillator system, valid in the regime of large oscillator excitation.
We follow the procedure of Ref.\onlinecite{AlsingCarmichael}, this time allowing for different qubit ($\omega_q$), oscillator ($\omega_p$), and drive ($\omega$) frequencies.
Our aim will be to find a closed set of equations for the time dependence of the expectation values of the oscillator and qubit coordinates, $\avg{a}$, $\avg{\sigma^-}$, and $\avg{\sigma^z}$.
We evaluate the time derivatives of these observables using $\frac{d}{dt}\avg{\mathcal{O}} = {\rm Tr}[\dot{\rho}\mathcal{O}]$, with the time-dependence of the density matrix given by Eq.(4) of the main text (below we set $\hbar = 1$),
\begin{equation}
\dot \rho =  - i \left[ H_{\rm RWA} , \rho\right]
      +  (n_B + 1) \Gamma
	\left( a\rho\adag - \frac{1}{2}\{\adag a,\rho\}\right) \nonumber\\
 +
	n_B \Gamma
	\left( \adag\rho a - \frac{1}{2}\{a\adag,\rho\}\right),
	\label{eqn:applindblad_terms}
\end{equation}
with $n_B = 1/(e^{\omega_p/k_B T}-1)$ and
$	H_{\rm RWA} =
	\frac {\tilde\omega_{q}}{2}  \sigma_3 +
	g  (a\sigma_+ + a^\dag \sigma_-) +
	\tilde\omega_p a^\dag a +
	\lambda (a+a^\dag)$.
Here, in the rotating frame, the reduced frequencies are given by $\tilde{\omega}_i = \omega_i - \omega$.
The qubit-phonon coupling is denoted by $g$, and the strength of the external driving field is denoted by $\lambda$.

Using the commutation rule $aa^\dagger = a^\dagger a + 1$ and the cyclic property of the trace, ${\rm Tr}[AB] = {\rm Tr}[BA]$, we find:
\begin{eqnarray*}
\avg{\dot{a}} &=& \left(-i\tilde\omega_p - \Gamma/2\right)\avg{a} - i\lambda - ig\avg{\sigma^-}\\
\avg{\dot{\sigma}^-} &=& -i\tilde\omega_q\avg{\sigma^-} + ig\avg{a\sigma_3}\\
\avg{\dot{\sigma}_3} &=& -2ig\left(\avg{a\sigma^+} - \avg{a^\dagger\sigma^-}\right).
\end{eqnarray*}
We close this set of equations by neglecting correlated fluctuations between the qubit and oscillator degrees of freedom, factoring the averages as $\avg{a\sigma_3} \approx \avg{a}\avg{\sigma_3}$ and $\avg{a\sigma^+} \approx \avg{a}\avg{\sigma^+}$.
Using $\avg{\sigma^+} = \avg{\sigma^-}^*$ and $\avg{a^\dagger} = \avg{a}^*$, and as in Ref.\onlinecite{AlsingCarmichael} defining the complex variables $z = \avg{a}$ and $v = 2\avg{\sigma^-}$, and a real variable $m = \avg{\sigma_3}$, we obtain
\begin{eqnarray}
\label{eq:z}\dot{z} &=& -(i\tilde\omega_p + \Gamma/2)z - \frac{i}{2}gv - i\lambda\\
\label{eq:v}\dot{v} &=& -i\tilde\omega_qv + 2igmz\\
\label{eq:m}\dot{m} &=& -ig(zv^* - vz*).
\end{eqnarray}
In this representation, the real and imaginary parts of $z$ describe the oscillator coordinate and momentum, the complex variable $v$ describes the $x$ and $y$ Bloch vector components of the qubit state, and $m$ describes the qubit polarization.

The dynamics described by the nonlinear system in
Eqs.~\eqref{eq:z}--\eqref{eq:m}
can be quite complex.
Here we focus on steady state solutions, $\dot{z} = 0, \dot{v} = 0, \dot{m} = 0$.
Setting $\dot v = 0$ in Eq.~\eqref{eq:v}, and introducing over-bars to indicate steady state values, we obtain
\begin{equation}
\label{eq:barv}\bar{v} = \frac{2g}{\tilde\omega_q}\bar{m}\bar{z}.
\end{equation}
Note that $\dot{m} = 0$ is automatically satisfied under this
condition, see Eq.~\eqref{eq:m}.

Within the semiclassical description, and in the absence of decoherence acting directly on the spin, the variables $v$ and $m$ describe a vector of unit length, $|v|^2 + m^2 = 1$.
Using Eq.~\eqref{eq:barv} for $\bar{v}$, we find $\bar{m}^2 = \left(1 + \frac{4g^2}{\tilde\omega_q^2}|\bar{z}|^2\right)^{-1}$.
Taking the square root of both sides gives
\begin{equation}
\label{eq:barm} \bar{m}_\pm = \pm\left(1 + \frac{4g^2}{\tilde\omega_q^2}|\bar{z}|^2\right)^{-1/2},
\end{equation}
where the subscript $\pm$ indicates two branches of solutions to the square root.
Setting $\dot{z} = 0$ in Eq.~\eqref{eq:z}, and using 
Eqs.~\eqref{eq:barv} and \eqref{eq:barm}, 
we find that the oscillator amplitude in the steady state satisfies the relation
\begin{equation}
\label{eq:barz}\frac{\lambda^2}{|z_\pm|^2} = \left(\Gamma/2\right)^2 + \left(\tilde\omega_p \pm \frac{g^2}{\sqrt{\tilde\omega_q^2 + 4g^2|z_\pm|^2}}\right)^2.
\end{equation}

Self-consistent solutions to Eq.~\eqref{eq:barz} 
can easily be found numerically.
In many parameter regimes, multiple solutions exist due to the non-linearity introduced by the qubit-oscillator coupling.
As shown in Fig.3e of the main text, the branches of stable fixed points match well with the peaks in the phonon number distribution, indicating the utility of the semi-classical approach.

Along with the presence of multiple solutions, we expected the typical manifestations of multistable behavior, such as hysteresis and sharp instabilities.
The sensitivity to the steady state oscillator amplitude near instability points where stable steady-state solutions disappear may be useful for sensing applications.
Closely related behavior has already proved quite useful in providing a sensitive read-out mechanism for superconducting qubits.\cite{Reed}

\section{Appendix D:
Interpretation of results for charge-sensing-based detection}

As stated in the main text, the oscillatory motion of the CNT resonator
can be detected using a charge sensing scheme
in which the conductance of a mesoscopic conductor, such as a QD or quantum point contact, is modulated via capacitive coupling to the charged mechanical resonator.
At a given source-drain bias on the mesoscopic conductor,
the current depends on the displacement $u$ of the resonator due to their capacitive
coupling.
A nonlinear dependence of the current
on the resonator displacement is desired for displacement sensing,
$I(t) \approx I_0 + I_1 u(t) + I_2 u^2(t)$.
Such a dependence is present in, e.g., a QD tuned to the middle of a
Coulomb-blockade peak (corresponds to $I_1 = 0$ and $I_2 < 0$),
or to the onset  of a Coulomb-blockade peak ($I_1>0$ and $I_2>0$).
Here we show that under such conditions, and neglecting
detector back-action on the oscillator, the steady-state time-averaged
current $\avg{I}$ through the charge-sensing mesoscopic conductor
is sensitive to the steady-state
average number $\bar n$ of phonons in the oscillator.
Therefore, the results plotted in Fig. 3b, d, g, h of the main text can be interpreted
as being proportional to the measured signal $\avg{I} - I_0$
in the charge-sensing setup described above,
hence that setup would allow for the experimental confirmation
of the predicted features.

The steady-state time-averaged current $\avg{I}$  through the charge-sensing
conductor is given by
\begin{equation}
\label{eq:avgcurrent}
\avg{I} = \lim_{\tau \to \infty} \frac 1 \tau \int_0^\tau \!\!\!\!\! dt\,  I(t)
= I_0 + I_2 \lim_{\tau\to \infty} \frac 1 \tau \int_0^\tau \!\!\!\!\! dt\,  u^2(t) .
\end{equation}
The instantaneous square displacement is expressed with
the steady-state density matrix $\bar \rho$ and the
position operator $x(t)$, both represented in the rotating frame, as
\begin{equation}
\label{eq:displacement}
u^2(t) = {\rm Tr} \left\{\bar \rho
x^2(t)\right\}
= {\rm Tr} \left\{\bar \rho
\left[\frac {\ell_0}{\sqrt 2} (a e^{-i\omega t} + a^\dag e^{i\omega
t})\right]^2\right\}
\end{equation}
Substitution of this expression to Eq. \eqref{eq:avgcurrent} yields
\begin{equation}
\avg{I} - I_0 = I_2 \ell_0^2 \left( \bar n + \frac 1 2 \right),
\end{equation}
as stated above.




\begin{thebibliography}{99}

\bibitem{OConnell-groundstate}
A. D. O'Connell, M. Hofheinz, M. Ansmann, R. C. Bialczak, M. Lenander, 
Erik Lucero, M. Neeley, D. Sank, H. Wang, M. Weides, J. Wenner, 
J. M. Martinis, and A. N. Cleland,
Nature \textbf{464}, 697 (2010).

\bibitem{Peng06}
H. B. Peng, C. W. Chang, S. Aloni, T. D. Yuzvinsky, and A. Zettl, 
Phys. Rev. Lett. {\bf 97}, 087203 (2006).

\bibitem{Chen09}
C. Chen, S. Rosenblatt, K. I. Bolotin, W. Kalb, P. Kim, I. Kymissis, 
H. L. Stormer, T. F. Heinz, and James Hone, 
Nature Nanotechnology {\bf 4}, 861 (2009).


\bibitem{steele09}
  G. A. Steele, A. K. H\"uttel, B. Witkamp, M. Poot, H. B. Meerwaldt, L. P. Kouwenhoven and H. S. J. {van der Zant}, Science \textbf{325}, 1103 (2009).

\bibitem{Chiu08}
H.-Y. Chiu, P. Hung, H. W. Ch. Postma, and M. Bockrath,
Nano Lett. {\bf 8}, 4342 (2008).

\bibitem{Lassagne08}
B. Lassagne, D. Garcia-Sanchez, A. Aguasca, and A. Bachtold, Nano Lett. {\bf 8}, 3735 (2008).

\bibitem{Lassagne11}
B. Lassagne, D. Ugnati, and M. Respaud, Phys. Rev. Lett. {\bf 107}, 130801 (2011).


\bibitem{loss-divincenzo}
  D. Loss and D. P. DiVincenzo, Phys.\ Rev.\ A {\bf 57}, 120 (1998).

\bibitem{Petta}
J. R. Petta, A. C. Johnson, J. M. Taylor, E. A. Laird, A. Yacoby, M. D. Lukin, C. M. Marcus, M. P. Hanson, and A. C. Gossard
Science {\bf 309}, 2180 (2005).

\bibitem{trauzettel07}
  B. Trauzettel, D. V. Bulaev, D. Loss, and G. Burkard, Nature Phys.\ \textbf{3}, 192 (2007).
           
\bibitem{bulaev08}
  D. V. Bulaev, B. Trauzettel, and D. Loss, Phys.\ Rev.\ B  \textbf{77}, 235301 (2008).
                 
\bibitem{ps}
  P. R. Struck and G. Burkard, Phys.\ Rev.\ B \textbf{82}, 125401 (2010).

\bibitem{Cottet-fmqubit}
A. Cottet and T. Kontos, Phys. Rev. Lett. {\bf 105}, 160502 (2010).

\bibitem{Rugar-singlespindetection} 
  D. Rugar, R. Budakian, H. J. Mamin and B. W. Chui, Nature \textbf{430}, 329 (2004).

\bibitem{Rabl-strongcoupling}
  P. Rabl, P. Cappellaro, M. V. Gurudev Dutt, L. Jiang, J. R. Maze and M. D. Lukin, Phys. Rev. B \textbf{79},  041302(R) (2009).

\bibitem{Rabl-spinphonon}
  P. Rabl, S. J. Kolkowitz, F. H. L. Koppens, J. G. E. Harris, P. Zoller and M. D. Lukin, Nat. Phys. \textbf{6}, 602 (2010).

\bibitem{Ando}
T. Ando, J. Phys. Soc. Jpn. {\bf 69} 1757 (2000).

\bibitem{Jeong}
J. S. Jeong and H. W. Lee, Phys. Rev. B {\bf 80}, 075409 (2009).

\bibitem{Izumida}
W. Izumida, K. Sato, and R. Saito, J. Phys. Soc. Jpn. {\bf 78}, 074707 (2009). 

\bibitem{Klinovaja}
J. Klinovaja, M. J. Schmidt, B. Braunecker and D. Loss, 
Phys. Rev. Lett. {\bf 106}, 156809 (2011).

\bibitem{kuemmeth08}
  F. Kuemmeth, S. Ilani, D. C. Ralph, P. L. McEuen, Nature \textbf{452}, 448 (2008).

\bibitem{Jespersen-cntspinorbit}
T. S. Jespersen, K. Grove-Rasmussen, J. Paaske, K. Muraki, T. Fujisawa, J. Nygard,
and K. Flensberg, Nat. Phys. {\bf 7}, 348 (2011).

\bibitem{Palyi-valleyresonance}
A. P\'alyi and G. Burkard, Phys. Rev. Lett. {\bf 106}, 086801 (2011).

\bibitem{suppmat}
See Supplemental Material below.

\bibitem{Mahan} 
G. D. Mahan, Phys. Rev. B {\bf 65}, 235402 (2002).

\bibitem{mariani09}
E. Mariani and F. von Oppen, Phys.\ Rev.\ B \textbf{80}, 155411 (2009).

\bibitem{Rudner-deflectioncoupling}
 M. S. Rudner and E. I. Rashba, Phys.\ Rev.\ B \textbf{81}, 125426 (2010).

\bibitem{Flensberg-bentnanotubes}
  K. Flensberg and C. M. Marcus, Phys. Rev. B \textbf{81}, 195418 (2010).

\bibitem{Huttel-cntresonator}
  A. K. Huttel, G. A. Steele, B. Witkamp, M. Poot,
  L. P. Kouwenhoven and H. S. J. van der Zant, Nano
  Lett. \textbf{9}, 2547 (2009).

\bibitem{churchill09}
  H. O. H. Churchill, F. Kuemmeth, J. W. Harlow, A. J. Bestwick, E. I. Rashba, 
  K. Flensberg, C. H. Stwertka, T. Taychatanapat, S. K. Watson, 
  and C. M. Marcus,
  Phys.\ Rev.\ Lett.\ \textbf{102}, 166802 (2009).


           
\bibitem{Jaynes}
E. T. Jaynes and F. W. Cummings,
Proceedings of the IEEE {\bf 51}, 89 (1963). 

\bibitem{AlsingCarmichael}
P. Alsing and H. J. Carmichael, Quantum Opt. {\bf 3}, 13 (1991).

\bibitem{Girvin}
L. S. Bishop, E. Ginossar, and S. M. Girvin, Phys. Rev. Lett. {\bf 105}, 100505 (2010).
		
\bibitem{Reed}
M. D. Reed, L. DiCarlo, B. R. Johnson, L. Sun, D. I. Schuster, L. Frunzio, 
and R. J. Schoelkopf,
Phys. Rev. Lett. {\bf 105}, 173601 (2010).


\bibitem{Knobel-nems-set}
R. G. Knobel and A. N. Cleland, Nature \textbf{424}, 291 (2003).
		
\bibitem{Elzerman04}
See e.g.,
J. M. Elzerman, R. Hanson, L. H. Willems van Beveren, B. Witkamp, 
L. M. K. Vandersypen, and L. P. Kouwenhoven,
Nature {\bf 430}, 431 (2004).

\bibitem{Wallraff05}
A. Wallraff, D. I. Schuster, A. Blais, L. Frunzio, J. Majer, M. H. Devoret, S. M. Girvin,
and R. J. Schoelkopf, 
Phys. Rev. Lett. {\bf 95}, 060501 (2005).


\bibitem{Imamoglu}
A. Imamoglu, D. D. Awschalom, G. Burkard, D. P. DiVincenzo, D. Loss, M. Sherwin and A. Small, Phys. Rev. Lett. {\bf 83}, 4204 (1999).

 \bibitem{Koppens}
 F. H. L. Koppens, C. Buizert, K. J. Tielrooij, I. T. Vink, K. C. Nowack, T. Meunier, L. P. Kouwenhoven, and L. M. K. Vandersypen, Nature {\bf 442}, 766 (2006).


\bibitem{wegewijs}
C. Ohm, C. Stampfer, J. Splettstoesser, and M. Wegewijs, Appl. Phys. Lett. {\bf 100}, 143103 (2012).





\end{thebibliography}

\begin{thebibliography}{99}

\bibitem{Weiss}
	S. Weiss et al., Phys. Rev. B \textbf{82}, 165427 (2010).


\bibitem{Jeong2009}
J.-S. Jeong and H.-W. Lee, Phys.~Rev.~B {\bf 80},  075409  (2009).

\bibitem{Izumida2009}
W. Izumida, K. Sato, and R. Saito, J. Phys. Soc. Jpn. {\bf 78},  074707
  (2009).

\bibitem{Klinovaja2011}
J. Klinovaja, M.J. Schmidt, B. Braunecker, D. Loss, Phys. Rev. B \textbf{84}, 0854452 (2011).

\bibitem{Jespersen1}
T. S. Jespersen, K. Grove-Rasmussen, J. Paaske, K. Muraki, T. Fujisawa, J. Nygard,
and K. Flensberg, Nat. Phys. {\bf 7}, 348 (2011).

\bibitem{Jespersen2}
T. Sand Jespersen, K. Grove-Rasmussen, K. Flensberg, J. Paaske,  K. Muraki, T. Fujisawa, J. Nygard,
Physical Review Letters \textbf{107}, 186802 (2011)

\bibitem{Bulaev}
	D. V. Bulaev, B. Trauzettel, and D. Loss, Phys.\ Rev.\ B \textbf{77}, 235301 (2008).

\bibitem{Ando-impurity}
T. Ando and T. Nakanishi, J. Phys. Soc. Jpn. {\bf 67}, 1704 (1998).

\bibitem{Palyi-valleyresonance}
A. P\'alyi and G. Burkard, Phys. Rev. Lett. {\bf 106}, 086801 (2011).

\bibitem{Flensberg-bentnanotubes}
  K. Flensberg and C. M. Marcus, Phys. Rev. B \textbf{81}, 195418 (2010).

\bibitem{AlsingCarmichael}
P. Alsing and H. J. Carmichael, Quantum Opt. {\bf 3}, 13 (1991).

\bibitem{Reed}
M. D. Reed, L. DiCarlo, B. R. Johnson, L. Sun, D. I. Schuster, L. Frunzio,
and R. J. Schoelkopf,
Phys. Rev. Lett. {\bf 105}, 173601 (2010).

\end{thebibliography}
\end{document}